\documentclass{aa}  

\usepackage[varg]{txfonts}
\usepackage[]{natbib}
\bibpunct{(}{)}{;}{a}{}{,}
\usepackage{graphicx}

\usepackage{savesym}
\savesymbol{tablenum}
\usepackage[print-unity-mantissa = false]{siunitx}
\restoresymbol{SIX}{tablenum}

\DeclareSIUnit{\MJ}{\emph{M}_\emph{J}}

\usepackage{soul}
\usepackage[colorlinks=true,linkcolor=blue, citecolor=blue, filecolor=blue, urlcolor=blue]{hyperref}

\begin{document}

   \title{The Outcome of Collisions between Gaseous Clumps\\ formed by Disk Instability}

   \author{Yoav Matzkevich\inst{1\star}
            \and
            Christian Reinhardt\inst{1,2}\thanks{These authors contributed equally to this work}
            \and
            Thomas Meier\inst{1}
            \and
            Joachim Stadel\inst{1}
            \and
            Ravit Helled\inst{1}
          }

   \institute{Department of Astrophysics, University of Zurich, Winterthurerstrasse 190, CH-8057 Zurich, Switzerland
    \and
    Physics Institute, Space Research and Planetary Sciences, University of Bern, Sidlerstrasse 5, CH-3012 Bern, Switzerland
    }

   \date{Received XXX / Accepted YYY}
 
\abstract{The disk instability model is a promising pathway for giant planet formation in various conditions. At the moment, population synthesis models are used to investigate the outcomes of this theory, where a key ingredient of the disk population evolution are collisions of self-gravitating clumps formed by the disk instabilities. 
In this study, we explore the wide range of dynamics between the colliding clumps by performing state-of-the-art Smoothed Particle Hydrodynamics simulations with a hydrogen-helium mixture equation of state and investigate the parameter space of collisions between clumps of different ages, masses (1--10 Jupiter mass), various impact conditions (head-on to oblique collisions) and a range of relative velocities.
We find that the perfect merger assumption used in population synthesis models is rarely satisfied and that the outcomes of most of the collisions lead to erosion, disruption or a hit-and-run. We also show that in some cases collisions can initiate the dynamical collapse of the clump.
We conclude that population synthesis models should abandon the simplifying assumption of perfect merging. Relaxing this assumption will significantly affect the inferred population of planets resulting from  the disk instability model. }


   \authorrunning{Y. Matzkevich et al.}
   \titlerunning{The Outcome of Collisions between Gaseous Clumps formed by Disk Instability}
   \maketitle

\section{Introduction} \label{sec:intro}

The two leading models for the formation of gas giants are core accretion \citep{safronovEvolutionProtoplanetaryCloud1972,goldreichFormationPlanetesimals1973,pollackFormationGiantPlanets1996} and disk instability (DI) \citep{kuiperOriginSolarSystem1951,bossGiantPlanetFormation1997,mayerFormationGiantPlanets2002}. The former is considered to be the standard theory which suggests that  giant planets form as a result of core formation followed by accretion of gas (mostly hydrogen and helium) from the proto-planetary disk. 

While core accretion offers predictions that are consistent with many observations, it is challenged by the presence of massive giant planets around low-mass stars \citep{bakosHATS71bGiantPlanet2020,delamerTOI4201EarlyDwarf2024,moralesGiantExoplanetOrbiting2019} and the formation of giant planets at wide orbital separations \citep{wagnerDirectImagesSpectroscopy2023,currieImagesEmbeddedJovian2022,sozzettiDynamicalMassGJ2023}. These observations prompt a further inquiry of the DI model, which offers possible cures for the short comings of core accretion \citep{helledGiantPlanetFormation2014}.

The DI model suggests that when a massive protoplanetary disk experiences gravitational instabilities, it can fragment into self-gravitating clumps that contract further to become gas giant planets. For razor-thin axisymmetric disks, instabilities occur according to the Toomre criterion \citep{toomreGravitationalStabilityDisk1964}:

\begin{equation}
Q\equiv\frac{c_S\kappa}{\pi G\Sigma}\lesssim 1, 
\end{equation}

where $c_S$ is the sound speed, $\kappa$ the epicyclic frequency and $\Sigma$ the surface density. In general, disks have a vertical extension and experience non-axisymmetric perturbations which explains why disk become unstable already at $Q\lesssim1.7$ \citep{durisenGravitationalInstabilitiesGaseous2007}. The fragmentation of the disk occurs if it can cool rapidly, with the cooling time being comparable to the  orbital period \citep{osullivanHydrodynamicRadiationTransfer2005}. 

The resulting fragments are expected to have masses of a few times the Jovian mass \SI{}{\MJ} \citep{boleyClumpsOuterDisk2010,rogersFragmentationProtostellarDiscs2012,forganJeansMassFundamental2011}, although the exact mass distribution is unknown and is still being investigated \citep{dengFormationIntermediatemassPlanets2021,kubliCharacterizingFragmentationSubJovian2023}.
During the first phase of the clump's evolution, it contracts quasi-statically and hydrogen remains in molecular form. The clump is very extended with its radius being on the order of an astronomical unit (AU). The duration of this period, called the pre-collapse stage, is around $10^4-10^6$ years \citep{decampliStructureEvolutionIsolated1979} depending on the clump mass \citep[e.g.,][]{helledHeavyelementEnrichmentJupitermass2009,helledMetallicityMassiveProtoplanets2010}.
As the clumps contract, their central temperature begins to rise and, when it reaches about 2000 K, the molecular hydrogen dissociates resulting in a dynamical collapse \citep[e.g.,][]{helledJupiterFormation2007,helledCoreFormationGiant2008}. This second phase occurs on a dynamical timescale of a few years \citep{bodenheimerCalculationsEvolutionGiant1980} resulting in a denser proto-planet with a radius of a few Jovian radii $R_J$ which continues to contract and cool down. This is known as the post-collapse phase with an associated timescale of $10^9$ years \citep{helledGiantPlanetFormation2014}. \\

The DI model can be investigated using various computational methods. 
Ideally, one should simulate the entire proto-planetary disk in 3D using a hydrodynamics code which follows the clumps'  formation and evolution as well as the interactions of the clumps with the disk and each other. 
However, such simulations  typically use  a simplified prescription of the gas physics, e.g., equation of state, opacities and cooling, and therefore are limited to simulating the earliest stages of a clump's evolution (e.g., \citealt{mayerFragmentationGravitationallyUnstable2007, dengFormationIntermediatemassPlanets2021}). In addition, the thermodynamics of the clumps is modelled in a simplified manner.
Previous studies simulated the clump's evolution using the appropriate equations of states and opacities, however, these models typically assume that the clump is isolated from the disk (e.g. \citealt{helledMetallicityMassiveProtoplanets2010}, \citealt{vazanEVOLUTIONSURVIVALPROTOPLANETS2012}) and therefore no interactions between clumps are considered. 

In order to understand the planetary population in the DI model population synthesis models must be performed  \citep{forganDynamicalFateSelfgravitating2015,humphriesConstrainingInitialPlanetary2019,schibLinkInfallLocation2023}. Such models aim to simulate the formation and early evolution of each self-gravitating fragment and determine how interactions with the central star, the disk or other fragments affect its physical and orbital properties.
A weakness of such models is that simplifying assumptions must be made when it comes to modelling the clumps' structure and the relevant physics involved. 
For example, population synthesis 
cannot capture the complex dynamics of colliding bodies since the simulations drastically simplify the clump interactions by either neglecting them or by considering the clumps as point masses that perfectly merge when colliding.

\citet{forganPopulationSynthesisModel2018} have shown that gravitational clump-clump interactions play a key role in sculpting the planetary population and motivated the investigation of collisions. \citet{hallIdentifyingAnalysingProtostellar2017} performed hydrodynamics simulations of disk fragmentation and followed the formation and early evolution of clumps as well as their mutual interactions. Their results highlight the importance  of gravitational interactions between clumps due to their effect on the clumps' final orbits and internal structure. Indeed, clump-clump interactions can lead to a variety of outcomes such as mergers, fragmentation, tidal interactions or even total disruption.

In this paper we investigate the outcome of  collisions between clumps using state-of-the-art 3D simulations. Our paper is organised as follows. Section \ref{sec:methods} describes the initial conditions for the impacts, the equation of state used in the simulations and the collision outcomes classification. Our results are presented and discussed in Sect. \ref{sec:results}. A summary and conclusions are   presented in Sect. \ref{sec:conclusions}.

\section{Methods} \label{sec:methods}

The simulations are performed using the Smoothed Particle Hydrodynamics (SPH)  method \citep{monaghanSmoothedParticleHydrodynamics1992}. SPH is well suited to model collisions and was applied to different impact regimes, ranging from Asteroids \citep{raducanGlobalscaleReshapingResurfacing2022, jutziConstrainingSurfaceProperties2022}, terrestrial \citep{kegerreisAtmosphericErosionGiant2020, timpeSystematicSurveyMoonforming2023,chauFormingMercuryGiant2018} to giant planets \citep{reinhardtBifurcationHistoryUranus2020, kegerreisConsequencesGiantImpacts2018, wooDidUranusRegular2022}. 
It exhibits great conservation properties, can handle geometries with high levels of deformation and tracks the fate of the material due to its Lagrangian nature. 
Our work constitutes the first application of SPH to the collision regime of proto-planetary clumps. For this study, we use a novel HPC implementation (Meier et al., in prep) of SPH  within the gravity code \texttt{pkdgrav3} \citep{potterPKDGRAV3TrillionParticle2017}. This SPH code leverages \texttt{pkdgrav3}, particularly its speed of gravity calculation on GPU/CPU parallel computers, while being a general SPH implementation with improvements for planetary collision simulations \citep{reinhardtNumericalAspectsGiant2017, reinhardtBifurcationHistoryUranus2020, meierEOSResolutionConspiracy2021, ruiz-bonillaDealingDensityDiscontinuities2022}.

\subsection{Initial conditions for the collisions} \label{subsec:ic}

The initial conditions for the collision simulations comprise of the pre-impact models of the two clumps (initial density, temperature, pressure profiles), the initial separation, the relative velocity at which they collide and the impact angle.

The pre-impact models of the clumps are taken from \citet{helledMetallicityMassiveProtoplanets2010} who modelled the pre-collapse  evolution of clumps using 1D hydrodynamic simulations. They considered  different clump masses ranging from 1 to $\SI{10}{\MJ}$. These models provide radial density, temperature and internal energy profiles of each clump. We consider four different clump masses of  1, 3, 5 and $\SI{10}{\MJ}$. For each mass we then consider three different models  that correspond to different evolutionary stages: 
the very extended stage, i.e., early evolution (henceforth called ``young"), right before dynamical collapse (``evolved") and at the median time between those two stages (``mid").

To obtain a low-noise particle representation of the models, we use the \texttt{ballic} code \citep{reinhardtNumericalAspectsGiant2017}. 
The particle representation is then evolved in isolation with \texttt{pkdgrav3} until it reaches equilibrium. Especially for young clumps the relaxation process requires adding a damper term due to their large extension and low density making them more sensitive to noise.

The number of particles used to represent each clump is chosen so that all particles in the simulation have the same mass to ensure numerical stability. We resolve the lowest mass, that is the $\SI{1}{\MJ}$ clump, with \SI{1e5}{} particles meaning that all the other clumps will have a higher resolution of \SI{3e5}{}, \SI{5e5}{} and \SI{1e6}{} for clumps with masses of $\SI{3}{}$, $\SI{5}{}$ and $\SI{10}{\MJ}$,  respectively. 
Using this resolution allows for the investigation of a large parameter space with a high reliability with  reasonable computational resources.

The initial conditions are set up such that the collision occurs in the xy-plane with the clump velocities being parallel to the x-axis at impact. 
In order to account for tidal deformation before the collision the clumps start with an initial separation of $5 R_{crit}$ where $R_{crit} = R_1 + R_2$ is the sum of the radii of both clumps
(see Figure~\ref{fig:ic} in the appendix for details). 

The relative velocities at impact  $v_{imp}$ that we investigate are normalised to the mutual escape velocity, defined as:

\begin{equation}
    v_{esc} = \sqrt{\frac{2G(M_1+M_2)}{R_1+R_2}}
\end{equation}

with $M_1,M_2$ and $R_1,R_2$ being the masses and radii of the colliding clumps. For the majority of our simulations, we consider  the interval $1 < v_{imp}/v_{esc} \leq 2$. This corresponds to impact velocities of a few \si{\kilo\meter\per\second} which is comparable to the orbital velocity at 100 AU assuming a solar mass star. Note that the possible range of relative velocities between two clumps orbiting a central star depends on the distance from the star and the eccentricity and inclination of the clump’s orbits. 
Collisions between clumps are expected to be above the mutual escape velocity even at large radial separations.
We use the impact parameter $b$ when the two clumps are in contact to parameterise the alignment of the collision. It is related to the impact angle by $b=\sin\theta$. A head-on collision corresponds to $b=0$ and a grazing collision to $b=1$, where the clumps only interact through tidal forces. In this study, we investigate collisions with $0\leq b\leq 0.7$. 

\subsection{Equation of state} \label{subsec:eos}

The adopted equation state (EoS) determines the type of material that is simulated. In this study, we use a realistic EoS modelling a mixture of the two most abundant gases in the proto-planetary disk, Hydrogen (H) and Helium (He). Initially developed by \citet{saumonEquationStateLowMass1995} for low-mass stars, brown dwarfs and giant planets, the SCvH (Saumon, Chabrier, van Horn) EoS models pure H, pure He and H-He mixtures with their non-ideal effects including ionisation, dissociation and formation of molecules. We will show in Sect.~\ref{sec:results} that these features are crucial for modelling the dynamics of clump collisions. The EoS was extended to lower pressures and temperatures as described in \citet{vazanEffectCompositionEvolution2013} to cover the conditions for the study of clumps since they tend to be  very diffuse and cold. The H-He mixture with a proto-solar composition of exactly 72.2\% H and 27.8\% He \citep{asplundChemicalCompositionSun2009} is calculated using the additive-volume rule.

\subsection{Collision outcomes} \label{subsec:outcomes}

In order to find gravitationally bound clumps after the collisions and to distinguish them from the surrounding ejecta, we use the group finder \texttt{skid} \citep{n-bodyshopSKIDFindingGravitationally2011} with parameters \textit{nSmooth} $=3200$ and \textit{tau} $=0.06$. The simulations are run until the number of clumps and the bound mass has converged.

We then quantify the outcomes through the ratio of final total bound mass to initial total mass $M_{tot}$ combined with the number of remaining fragments $N_{frag}$ after the collision. The classification of the collision outcomes is summarised in Table~\ref{tab:collision_outcomes}.  In some cases, the fragments can survive a first collision but remain gravitationally bound and re-collide later to merge. These so called graze-and-merge collisions form a sub-regime of mergers and are at the transition to the hit-and-run regime \citep{stewartCollisionsGravitydominatedBodies2012, chauCouldUranusNeptune2021} but can require more simulation time to be resolved \citep{timpeSystematicSurveyMoonforming2023}.

\begin{table}
\caption{Collision outcomes\label{tab:collision_outcomes}}
\centering
\begin{tabular}{lccc}
\hline\hline
Description & $N_{frag}$ & $M_{bound}/M_{tot}$\\
\hline
Perfect merger & 1 & $\geqslant$ 0.95 \\
Erosion & 1 &  $\geqslant$ 0.5 \\
Disruption & 0/1 & $<$ 0.5 \\ \hline
Perfect hit-and-run & 2 &  $\geqslant$ 0.95 \\
Erosive hit-and-run & 2 & $<$ 0.95\\
\hline
\end{tabular}
\tablefoot{Classification of the collisions depending on the final number of bound fragments as well as ratio of final total bound mass to initial total mass. By definition, a perfect merger is a collision where 100\% of the colliding mass remains bound in a single fragment. To relax this strict requirement, we consider a perfect merging as a collision with a mass loss of up to 5\%. As discussed in Sect.~\ref{sec:intro}, we expect diverse outcomes from the rich dynamics of the collisions.}
\end{table}

\section{Results and discussion} \label{sec:results}

In total, we performed over two hundred simulations whose results are organised here according to the classification presented below and in Table~\ref{tab:collision_outcomes}. 
A summary of all the simulations with their initial conditions is listed in Table~\ref{tab:sim_database}. Table~\ref{tab:outcome_database} lists the occurrence of each type of outcome depending on the pair of colliding clumps. The collision outcomes depend on the clump mass and age as well as the impact parameter and velocity. The parameter space we consider here captures the transition between a large diversity of post-collision states. 

\subsection{Mergers}\label{subsec:mergers}
We define a merger as a collision between two clumps resulting in a single gravitationally bound object. In a perfect merger (PM), as assumed in population synthesis models, this object contains the sum of their masses.
However, in our set of simulations PM are very rare even for the most favourable conditions, that is for evolved (and therefore more compact) clumps and oblique impacts at $v_{imp} \sim v_{esc}$. Of all the 40 impact simulations between evolved clumps of different masses we find that only four result in PM, two of them being graze-and-merge collisions (for details see Sect.~\ref{subsec:outcomes}). 

\begin{figure}[ht!]
\centering
\includegraphics[width=\hsize]{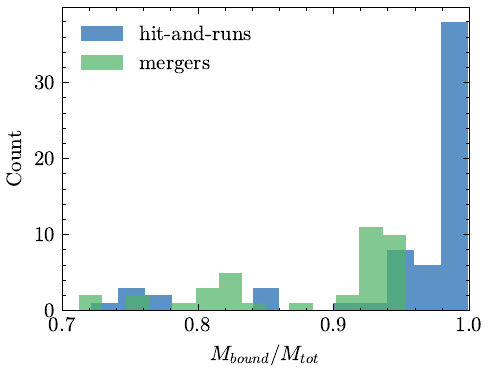}
\caption{Distribution of final bound mass fraction for post-collision states in the case of a merger (see Sect.~\ref{subsec:mergers}) or a hit-and-run (see Sect.~\ref{subsec:HRC}). The majority of mergers are erosive while there are mainly perfect rather than erosive hit-and-runs. Statistics for disruptions are shown in Fig.~\ref{fig:distrib_disrupt_HR}.
\label{fig:mass_hist}}
\end{figure}

Figure \ref{fig:mass_hist} shows the distribution of the post-impact bound mass for mergers and combined mass of the two fragments for hit-and-run collisions (HRC), in both cases the mass is normalised to the initial total mass in the collision. The majority of observed mergers are erosive with more than 5\% of mass loss after the collision and therefore are not classified as PM. 

\begin{figure*}
\centering
\includegraphics[width=\hsize]{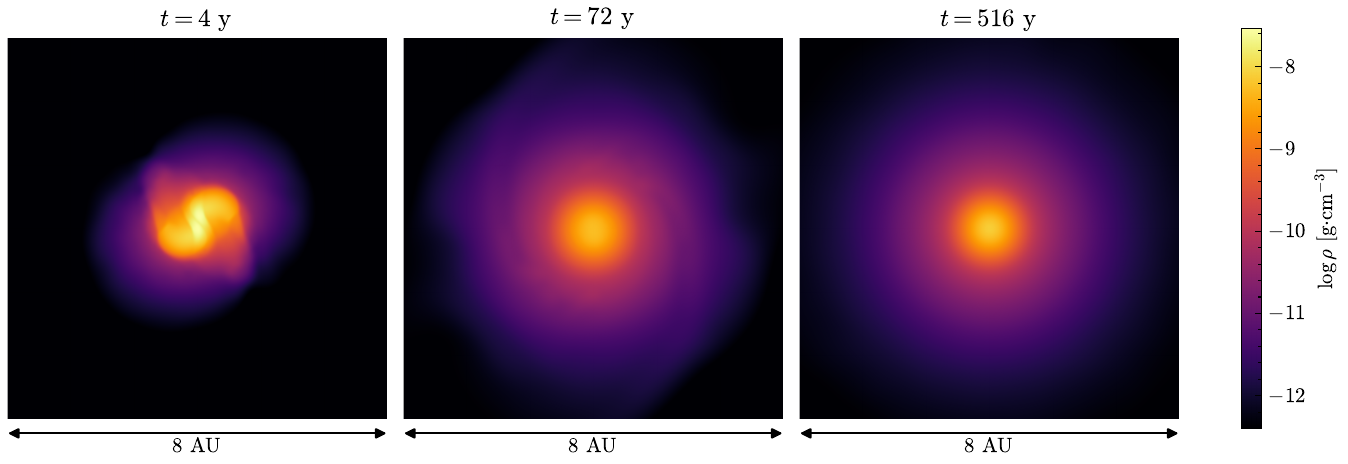}
\caption{Snapshots of slices in the z-direction from the simulation of an erosive merger between two evolved \SI{5}{\MJ} clumps colliding with $b = 0.3$ and $v_{imp}/v_{esc} = 1.5$. Since the collision is oblique, spiral structures are emerging but are then falling back onto the bound remnant until it reaches a spherically symmetric equilibrium state containing 72 \% of the initial total mass. 
\label{fig:merger}}
\end{figure*}

Figure~\ref{fig:merger} shows an example of an erosive merger between two evolved \SI{5}{\MJ} clumps colliding with $b = 0.3$ and $v_{imp}/v_{esc} = 1.5$. The snapshots are slices in the z-direction and show how the clumps collide, merge, and finally reach a hydro-static equilibrium. 
An important aspect of those collisions is their large scale compared to planetary collisions. This example shows how the clumps collide and spread out material on distances in the order of the AU.
For such extended objects the point-mass approximation used in population synthesis models is inappropriate. 

\subsection{Hit-and-run collisions}\label{subsec:HRC}
\begin{figure*}
\sidecaption
\includegraphics[width=12cm]{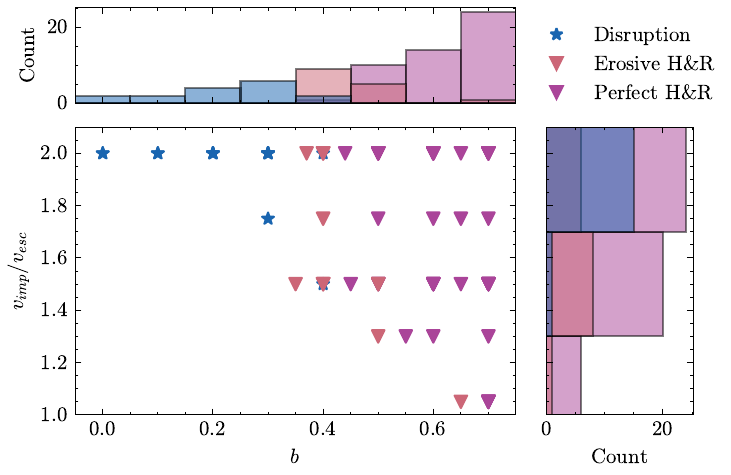}
\caption{Distribution of all observed erosive/perfect hit-and-runs (see Sect.~\ref{subsec:HRC}) and disruptive collisions (see Sect.~\ref{subsec:disruption}) depending on impact parameter and impact velocity. Hit-and-runs are favoured by oblique collisions and at the transition to the other regimes, they become erosive. 
Disruptions are favoured by high impact velocities and head-on collisions.
\label{fig:distrib_disrupt_HR}}
\end{figure*}

For impact velocities above the mutual escape velocity, oblique collisions result in hit-and-run collisions (HRC) where both clumps 
survive the encounter with a deviated trajectory after exchanging mass and angular momentum.
Figure~\ref{fig:distrib_disrupt_HR} shows the distribution of all the observed HRC depending on $b$ and $v_{imp}/v_{esc}$. 
The majority of such events are preserving the mass of the clumps as shown in Fig. \ref{fig:mass_hist}. Erosive HRC, where part of the bodies mass is lost due to the impact, are less common. 

This is a result of the large radial density gradients inside the clumps. Most of their mass is contained within their inner layers, and therefore erosion of the outer layers by grazing collisions is less significant. The magnitude of the erosion strongly  depends on the clump's age since young clumps have lower central densities compared to evolved clumps, making them prone to lose more material during an HRC. 

\begin{figure*}[ht!]
\centering
\includegraphics[width=\hsize]{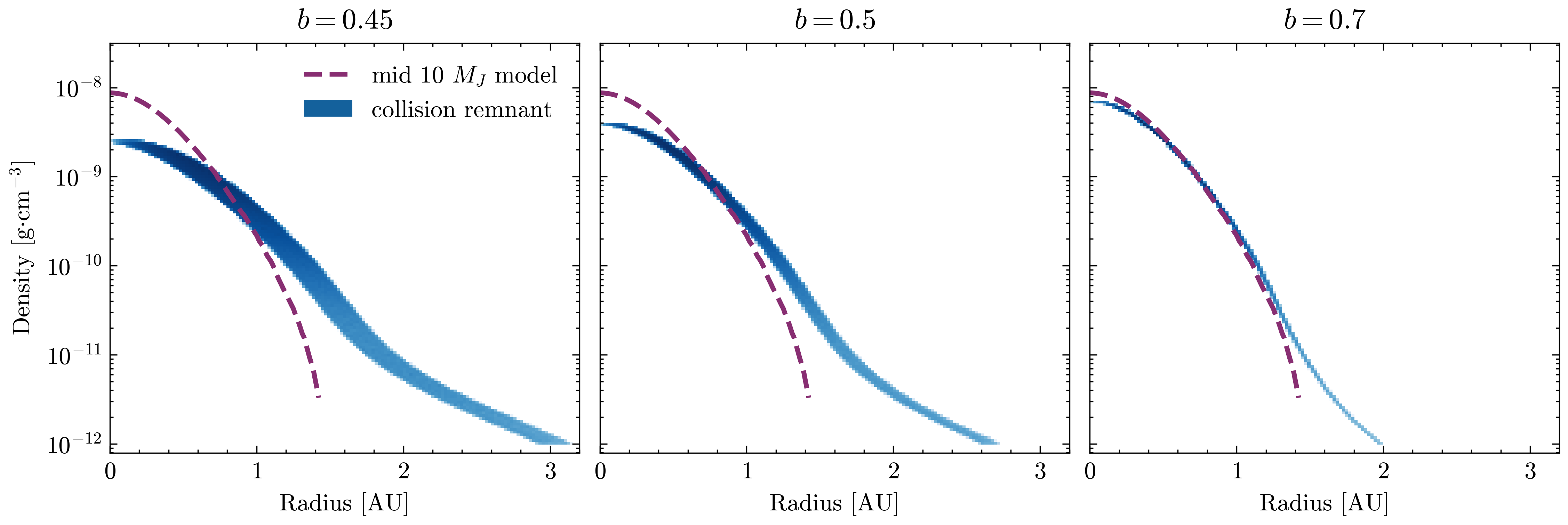}
\caption{Radial profiles for the density of hit-and-run remnants from $v_{imp}/v_{esc} = 1.5$ collisions between two mid \SI{10}{\MJ} clumps (the two remnants have the same profile due to symmetry). Only the gas with a density higher than \SI{1e-12}{\gram\per\centi\meter\tothe3} is displayed. The profile of the mid \SI{10}{\MJ} model is shown for reference. For each impact parameter the post-collision bound mass to total mass ratio is respectively 95.3 \%, 97.5 \% and 99.7 \% with the angular momentum of the remnant being 2.7\%, 2\% and 0.7\%  of the initial angular momentum of the collision, respectively. Head-on collisions essentially perturb the structure of the colliding clumps more than oblique collisions.}
\label{fig:HR_b}
\end{figure*}

We observe that a change of the impact parameter
affects the core's density: the more oblique the collision, the higher the inner density of the remnants. A larger impact parameter results in remnants resembling the pre-collision clump profiles with only the outer layers being perturbed. More head-on collisions allow a greater exchange of kinetic energy and angular momentum. This leads to the spreading of material outwards due to the induced rotation and shock heating. 
This also affects the inner structure of the clump and can reduce the core density by almost an order of magnitude. Figure \ref{fig:HR_b} illustrates this phenomenon by showing the radial density profiles for remnants of collisions between two mid \SI{10}{\MJ} clumps at $v_{imp}/v_{esc} = 1.5$ with the exact same initial conditions except a varying impact parameter. The post-collision angular momentum of the clumps is twice as large for $b=0.5$ than for $b=0.7$  which corresponds to 2\% and 0.7\% of the respective initial orbital angular momentum of the system. 

While HRC are very common during the formation of rocky planets \citep[e.g.,][]{asphaugHitandrunPlanetaryCollisions2006,kokuboFORMATIONTERRESTRIALPLANETS2010}, 
it remains unclear whether such events are common for gaseous clumps. In case they occur as frequently as for rocky planets, HRC need to be considered in population synthesis models since they deviate the trajectory of the clumps and reduce their kinetic energy. 
For clumps, HRC preserve the bound mass better than mergers but can still affect the stability of the clump in future encounters since a higher core density allows the clump to be sufficiently gravitationally bound to survive further collisions.

\subsection{Disruption}
\label{subsec:disruption}

Collisions with high impact velocities and low impact angles can lead to significant mass loss or even destruction of the clumps. 
If more than half of the initial total mass is lost, we refer to the outcome of the collision  as ``disruption". 
Such disruptions occur when the shock wave accelerates a significant amount of bound material beyond the escape velocity, ejecting it out of the clump's gravitational potential. Clumps tend to be weakly bound, especially when they are young and are not very massive, i.e., the 1 and \SI{3}{\MJ} clumps in our simulations. As a result,  their escape velocity is low and disruptive collisions are much more common than for compact and dense objects like  rocky planets. In our set of collisions, we observe that these outcomes are favoured by large impact velocities and frontal collisions as shown in Fig. \ref{fig:distrib_disrupt_HR}. 

\begin{figure*}[ht!]
\centering
\includegraphics[width=\hsize]{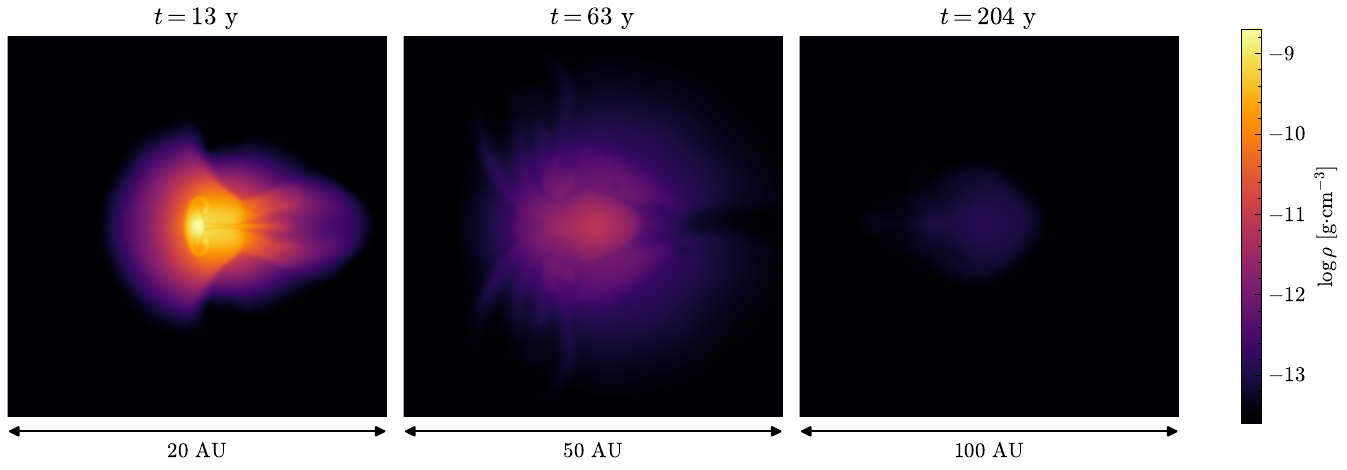}
\caption{Snapshots of slices in the z-direction from the simulation of a super-critical disruption between mid 10 and \SI{5}{\MJ} clumps with $b = 0$ and $v_{imp}/v_{esc} = 2$. The shock wave propagates in both clumps and accelerates the gas beyond the mutual escape velocity. The ejected material is very low in density and is distributed over $\sim 100$ AU (note the different length scale of each frame). This event clearly shows how the population of clumps can decrease as a result of a collision. 
}
\label{fig:disruption}
\end{figure*}

Figure \ref{fig:disruption} shows an example of a collision between a mid \SI{10}{\MJ} and a mid \SI{5}{\MJ} clumps with $b = 0$ and $v_{imp}/v_{esc} = 2$ that leads to a super-critical disruption. The strong shock wave caused by the collision propagates in a symmetric way since the collision is head-on. The material  is ejected and no bound mass is left after the collision. 

Disruption is more common with younger clumps since they are more extended and are less gravitationally bound: while the transition to disruption for evolved clumps is at $v_{imp}/v_{esc} \simeq 2$ (corresponding to velocities in the order of 7--9 \SI{}{\kilo\meter\per\second}), young clumps can already be disrupted by collisions at $v_{imp}/v_{esc} \simeq 1.5$ (in the order of 2--3 \SI{}{\kilo\meter\per\second}). Such velocities are comparable to the Keplerian velocity at 80 AU around a solar mass star. Therefore, disruptive outcomes could be very common during the early evolution of clumps and should be considered in population synthesis models. 

\subsection{Core collapse}\label{subsec:collapse}

Some of the simulations, consisting of mid and evolved clumps involved in frontal collisions exhibit a fast and intense increase of density in distinct regions where they reach \SI{1}{\gram\per\centi\meter\tothe3} compared to the maximal value of about \SI{1e-6}{\gram\per\centi\meter\tothe3} reached by the regular collisions. The time span needed to reach those high densities is in the order of a few years, which is much shorter than the entire collision timescale.\footnote{These simulations grind to a halt at this time. Modelling the dynamic collapse requires very small time steps and therefore such simulations are computationally very expensive.} 

\begin{figure*}
\sidecaption
\includegraphics[width=12cm]{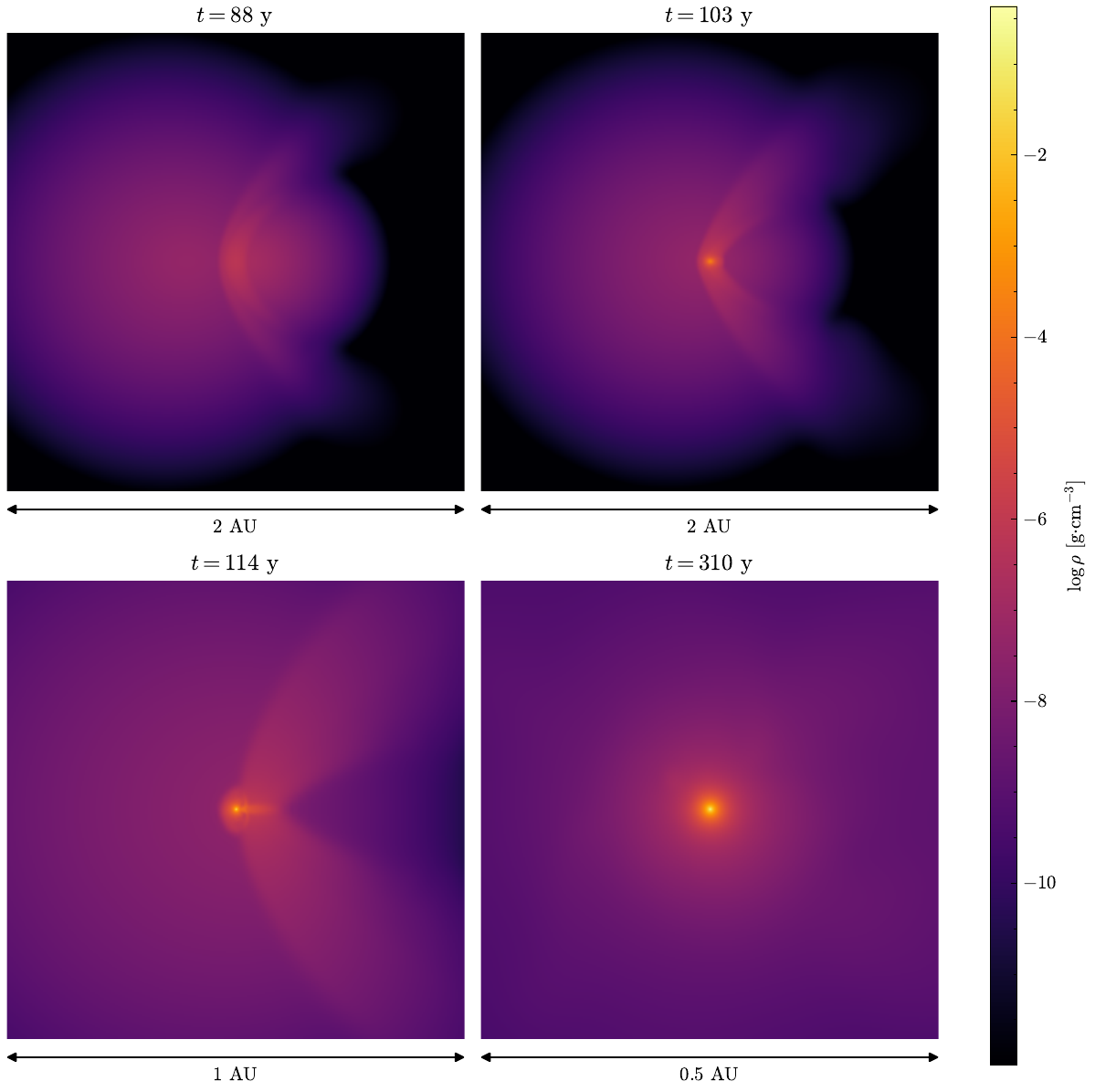}
\caption{Snapshots of slices in the z-direction from the simulation of a collapsing collision between evolved 10 and \SI{5}{\MJ} clumps with $b = 0$ and $v_{imp}/v_{esc} = 1.05$. 
During the impact a compact region is forming behind the shock front in which the density increases by several orders of magnitude (top panels).
At a later stage the collapsed region pierces through the shock front (bottom left panel) and regains hydrostatic equilibrium (bottom right panel).
After the collision \SI{77}{\percent} of the total colliding mass remains bound.
Such collisions are a pathway to shorten the pre-collapse stage and influence the survival of the clumps in the proto-planetary disk.}
\label{fig:collapse}
\end{figure*}

Figure \ref{fig:collapse} shows such an event between evolved 10 and \SI{5}{\MJ} clumps in a collision with $b = 0$ and $v_{imp}/v_{esc} = 1.05$. We see the formation of a collapsing region  behind the shock front that leads to a denser and more compact core. In this example, the bound mass of the remnant is \SI{12.04}{\MJ} which corresponds to  $\sim$77\% of the initial total mass.

By compressing the gas during the collision, its pressure, density and temperature increase. The entropy also increases due to shock heating 
but in the case of adiabatic decompression such as during shock release, the evolution is isentropic. While the isentropes in the low pressure and temperature range of the SCvH table are identical to those of an ideal gas with a suitable molecular mass, there is a change at around $\log T \simeq 3$. From that point, all isentropic curves exhibit a relatively flat region in the sense that the density increases by several orders of magnitude without a significant increase in temperature.

This  kink in the isentropes corresponds to a phase transition since at these temperatures the molecular hydrogen starts to dissociate
\citep{saumonEquationStateLowMass1995}. The kinetic energy gained from the collision is converted into energy for breaking the molecular bonds of the gas. Since the thermal pressure cannot balance gravity at these augmented densities, the clump is no longer  in hydrostatic  equilibrium leading to a collapse that occurs on a timescale of the order of the free-fall time $t_{f\!f}\sim1/\sqrt{G\rho}$ \citep[e.g.,][]{bodenheimerCalculationsEvolutionGiant1980,helledHeavyelementEnrichmentJupitermass2009,helledMetallicityMassiveProtoplanets2010}. By calculating $t_{f\!f}$ with the mean density of the clumps we indeed find that it is $\sim 2$ years, which is consistent with our simulations that were run through the collapse phase.

\begin{figure*}[ht!]
\centering
\includegraphics[width=\hsize]{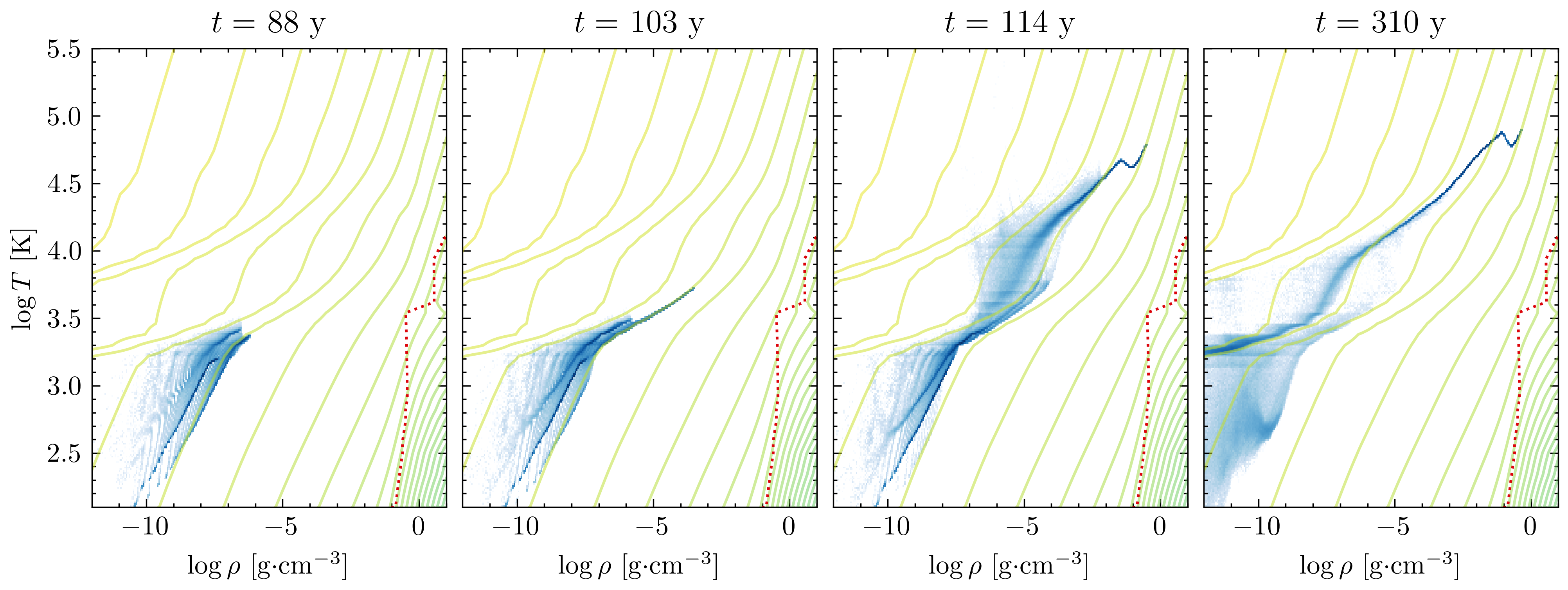}
\caption{Evolution of the SPH particles from Fig. \ref{fig:collapse} in the logarithmic $\rho-T$ plane, in which the curves are $(\rho, T)$ at constant entropy and the flat regions are phase transitions, namely dissociation of molecular hydrogen.
Dotted red lines indicate where the original EoS table was extrapolated and the thermodynamic quantities can exhibit unphysical behaviour.
The structures of the two colliding clumps can be distinguished from each other in the first frames but are then mixed by the impact. }
\label{fig:collapse_scvh}
\end{figure*}

To investigate the collapse stage, the SPH particles of the same simulation snapshots from Fig. \ref{fig:collapse} are shown 
in the $\rho-T$ diagram with curves of constant entropy in Fig. \ref{fig:collapse_scvh}. The first frame shows how, during the collision, the particles are compressed and entropy increases due to shock heating.
Once particles enter the flat region corresponding to molecular hydrogen dissociation, the dynamical collapse is initiated.
After the collision, most of them reach the upper-right region of the diagram and stay with these high densities and temperatures: these are the ones making up the core. The innermost part of the core stays on its isentrope while the rest of the clump has been shock heated during the rapid in fall.

The simulations show that the collapsing region is behind the shock wave
due to the lower thermal pressure that acts against gravity. Indeed, within the shock front the evolution is non-adiabatic (see Sect.~\ref{sec:methods}) and gas heating provides support against collapse.
Nevertheless, the collapsed region manages to pierce through the shock front after having formed and then accretes more mass due to its high density.

\begin{figure*}[ht!]
\centering
\includegraphics[width=\hsize]{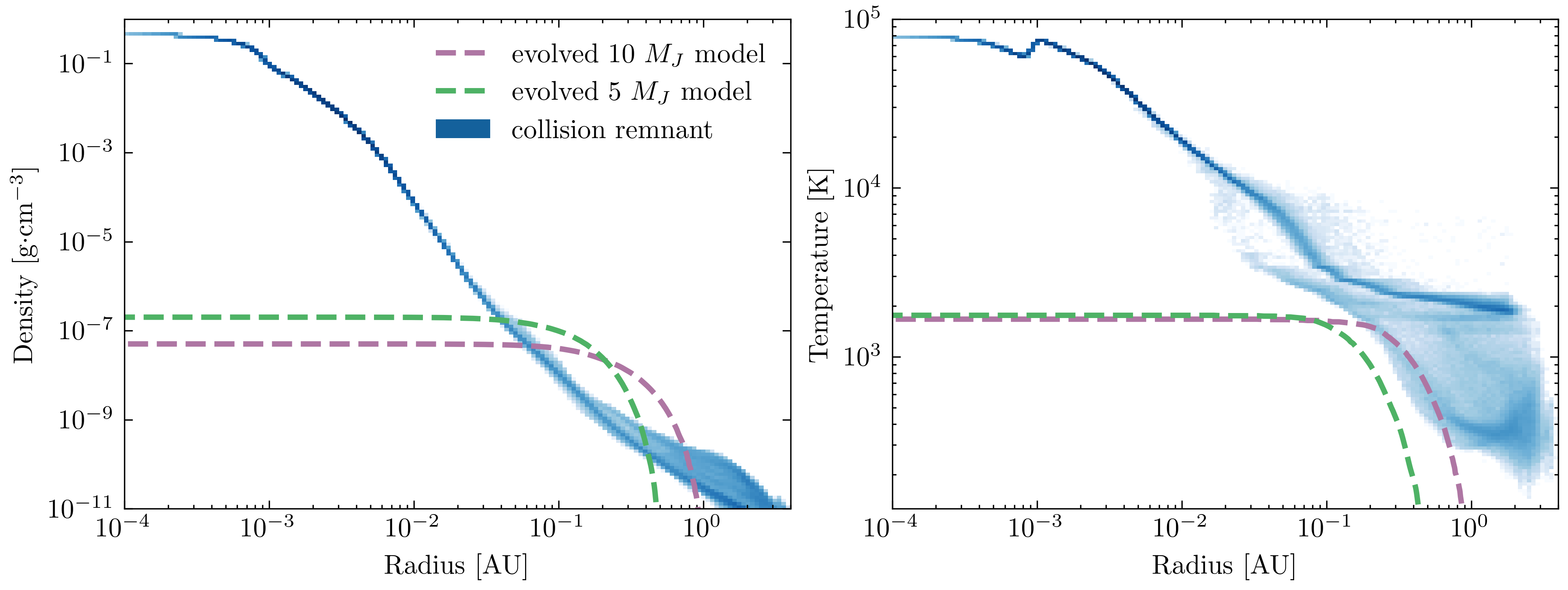}
\caption{Radial density and temperature profiles for the SPH particles of the remnant resulting from the collapsing collision displayed in Fig. \ref{fig:collapse}. Only the particles with a density higher than \SI{1e-11}{\gram\per\centi\meter\tothe3} are shown. The bound mass of the remnant is \SI{12.04}{\MJ} amounting to 77\% of the initial total mass.
The densities and temperatures of the collision remnant are extremely high compared to the the initial evolved 10 and \SI{5}{\MJ} clump models, whose profiles are shown for reference. 
A kink in the temperature at $R=\SI{1e-3}{}$ AU is due to shock heating of material falling onto the central dense core.}
\label{fig:collapse_remnant}
\end{figure*}

After the collapse, the remnant evolves towards hydrostatic equilibrium
and regains spherical symmetry. Figure \ref{fig:collapse_remnant} compares the radial density and temperature profiles of the remnant
to the initial models of the two colliding clumps. The differences in temperature and density are significant. The central part with a radius of a few $R_J$, is almost at $\rho=\SI{1}{\gram\per\centi\meter\tothe3}$ which corresponds to Jupiter's mean density and the temperatures at around 70,000 K, comparable to central temperatures expected from the core accretion model  \citep{mullerChallengeFormingFuzzy2020}. In principle, some energy should be lost via radiative cooling which is not considered in the simulations. As a result, our inferred temperature should be taken as upper bounds.

In order to investigate the physical nature of this outcome, the exact same collisions were simulated using an ideal gas EoS. These simulations do not result in dynamical collapse but are classified either as erosive mergers or, for high impact velocities, as disruptions, as defined in Table \ref{tab:collision_outcomes}.
This highlights the importance of using appropriate EoSs in impact simulations. 
As another sanity check, we performed a resolution study on a core collapse event between two evolved \SI{10}{\MJ} colliding with $b=0$ and $v_{imp}/v_{esc}=1.05$ to check whether this phenomenon is resolution dependent. The simulations with $N=10^5$, $10^6$ and $10^7$ particles all experience the same dynamical collapse around the same time step which further supports that it is not a numerical artefact but indeed a physical event being captured by the SPH code.

More than 25\% of the simulations lead to a dynamical collapse and it is more common for simulations with evolved clumps.  
That is because younger clumps require more energy to reach molecular hydrogen dissociation but are also more easily disrupted by strong shocks. On the other hand, evolved clumps are denser and hotter to begin with and therefore a collision can trigger the molecular dissociation for a wider range of impact parameters and velocities. Evolved clumps are also more stable against disruption due to their compactness and therefore the kinetic energy of the collision can efficiently contribute to the collapse without destroying the clump. 
Earlier collapse can also be initiated for collisions between clumps that are in the middle of the pre-collapse evolution (see Sect.~\ref{subsec:overview}). Even collisions of mixed ages can trigger the dynamical collapse within the clumps, as we observe for instance in those between young \SI{3}{\MJ} and evolved \SI{10}{\MJ} clumps, where the collision is triggered within the \SI{10}{\MJ} clump. 

The initiation of a dynamical collapse via collisions shortens the pre-collapse stage of clumps. This would affect the inferred population since the remnants resulting from the dynamical collapse have higher chances to survive. As a result, this mechanism should be considered by  population  synthesis models. 

\subsection{Overview}\label{subsec:overview}

In this section we summarise the outcomes of the different collision  simulations.
For different combinations of clump mass and age we plot the outcomes defined in Table \ref{tab:collision_outcomes}, as well as the core collapse, on a diagram with $b$ and $v_{imp}$ as coordinates allowing for an overview of the results.

\begin{figure}[ht!]
\centering
\includegraphics[width=\hsize]{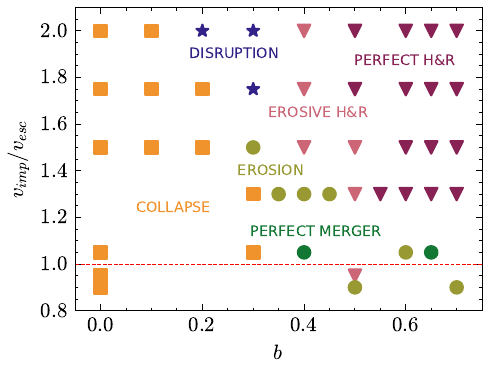}
\caption{Collision outcomes from simulations of two  evolved \SI{10}{\MJ} clumps. The markers represent the real data acquired from the simulations. 
The data point at $(b,v_{imp})=(0.65,1.05)$ is a case of graze-and-merge impact where $\geq 95\%$ of the mass remains gravitationally bound. This plot includes additional simulations at impact velocities below the mutual escape velocity (dashed red line). Head-on collisions with $v_{imp}/v_{esc} < 1$ still trigger dynamical collapse while more oblique impacts result in erosion or hit-and-run.}
\label{fig:overview_old}
\end{figure}

Figure \ref{fig:overview_old} shows such a diagram for impacts between two evolved \SI{10}{\MJ} clumps. The diagram for two colliding evolved \SI{5}{\MJ} clumps looks exactly the same suggesting that the age and mass ratio is more relevant than the total  colliding mass. 
Evolved and massive clumps have higher chances to collapse as a result of a collision since they are hotter, denser and more resistant against disruption compared to young and low-mass clumps.  
If the impact velocity is too high, there is a risk of eroding or disrupting both clumps when the collision occurs at low impact parameters. For more oblique impacts both clumps survive hit-and-run collisions. Erosive or perfect mergers are observed only in a limited region of this parameter space. This means that if two evolved clumps collide, they are more likely to retain their initial masses and have higher chances to evolve into  giant planets as they can reach the dynamical collapse stage faster.  It is still unclear whether such collisions are more common for impacts below the escape velocity since they might also collapse (see discussion below). 

\begin{figure}[ht!]
\centering
\includegraphics[width=\hsize]{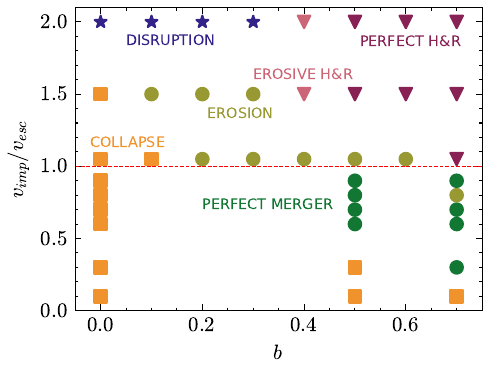}
\caption{Collision outcomes from simulations of two  mid 10 and \SI{5}{\MJ} clumps. There are no PM in this range of parameters and the collapse region becomes smaller compared to Fig. \ref{fig:overview_old}. Additional simulations at impact velocities below the mutual escape velocity (dashed red line) show that head-on collisions with $\sim$\SI{350}{\meter\per\second} initiate dynamic collapse. Only very grazing impacts can
result in simple PM (see discussion in Section~\ref{subsec:lowvel}).
}
\label{fig:overview_mid}
\end{figure}

Figure \ref{fig:overview_mid} shows the same diagram for collisions between mid \SI{10}{\MJ} and mid \SI{5}{\MJ} clumps. While the mass ratio is different from 1:1, the overall structure is similar to the outcomes of collisions between two evolved \SI{10}{\MJ} clumps. Since the mid clumps are younger, they are more extended and therefore more vulnerable for disruption. As a result, PM does not occur and the mergers that do take place are erosive. Head-on collisions lead to disruption more often and the region where collapse can be initiated is smaller compared to evolved clumps. If a collision can initiate dynamical collapse, the duration of the pre-collapse stage of mid clumps is halved. In the case of a \SI{5}{\MJ} clump, this corresponds to \SI{1.2e4}{} instead of \SI{2.4e4}{} years. This significantly improves the likelihood of surviving future encounters. 

\begin{figure}[ht!]
\centering
\includegraphics[width=\hsize]{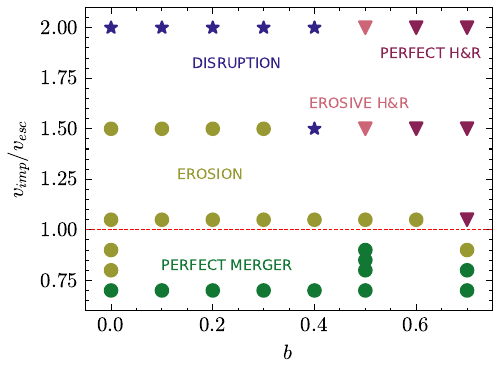}
\caption{Collision outcomes from simulations of two young \SI{3}{\MJ} clumps.  Such clumps are very extended and have low densities. Therefore, their gravitational binding energy is very low and impacts result in either erosion or disruption. PM or dynamic collapse are not  observed.
Additional simulations at impact velocities below the mutual escape velocity (dashed red line) show that head-on or very grazing ($b\sim0.7$) impacts with $v_{imp}/v_{esc} \geq 0.8-0.9$ corresponding to \SIrange{1.36}{1.54}{\kilo\meter\per\second} still lead to erosion. Only intermediate impact parameters result in PM (see discussion in Section~\ref{subsec:lowvel}).
}
\label{fig:overview_early}
\end{figure}

Figure \ref{fig:overview_early} shows the outcomes of two colliding young \SI{3}{\MJ} clumps. In this case collisions do not initial dynamical collapse. 
Young and less massive clumps are very fragile and easy to disrupt, however, they can still survive after HRC. We find no PM in this range of parameters. In fact, for a purely head-on collision PM occurs only for an impact velocity as low as $v_{imp}=0.7 v_{esc}$ which corresponds to a relative velocity of $\sim$1.2 \SI{}{\kilo\meter\per\second}. We therefore conclude that when young clumps collide the assumption of PM is inappropriate. 

\subsection{Low velocity impacts}\label{subsec:lowvel}
All the simulations presented above have impact velocities above the mutual escape velocity. Although low velocity collisions are expected to be less common, in this section we perform additional simulations of head-on ($b=0$) and oblique ($b=0.5,\ 0.7$) impacts for the clump models shown in Figs.~\ref{fig:overview_old}, \ref{fig:overview_mid} and \ref{fig:overview_early} with velocities lower than the mutual escape velocity.

The results of such collisions between two evolved \SI{10}{\MJ} clumps are shown in Fig.~\ref{fig:overview_old}. We find that head-on collisions can still trigger dynamical collapse and that  grazing impacts lead to erosion or erosive hit-and-run collisions. Even for impacts below the mutual escape velocity we find that perfect mergers are a relatively rare outcome which occurs only for very grazing and low velocity collisions. The outcomes of collisions between two mid \SI{10}{\MJ} and \SI{5}{\MJ} clumps are shown in Fig.~\ref{fig:overview_mid}. Such collisions can lead to a dynamical collapse even for very low impact velocities. Grazing impacts above the mutual escape velocity result in erosion of hit-and-run, and collisions with $v_{imp}/v_{esc} \geq 0.6$ for $b=0.5$ and $v_{imp}/v_{esc} \geq 0.3$ for $b=0.7$ can lead to perfect mergers. Finally, we find that collisions between two young \SI{3}{\MJ} clumps (Fig. \ref{fig:overview_early}) are still erosive slightly below the mutual escape velocity if they are head-on ($b=0$) or very grazing ($b=0.7$). 
For the other cases collisions can lead to PM. Overall, we conclude that even for collisions with  $v_{imp}/v_{esc}<1$ the assumption of PM is often inappropriate. It is yet to be determined how relevant low velocity and oblique collisions are for young clumps. Preliminary results from population synthesis models in the disk instability scenario (Schib et. al., in prep.) suggest that impacts below the mutual escape velocity are very rare and that head-on collisions are more common.

\subsection{Limitations and future work}\label{subsec:outlook}
While this study constitutes a first step in understanding the dynamics of colliding proto-planetary clumps, future studies should continue the investigation of this topic including more processes and a larger parameter space as we discuss below.

First, future research could obtain the impact conditions from 
disk simulations which follow the formation and evolution of clumps formed  by DI. This could provide a more realistic parameter space for studying collisions in terms of clump sizes and masses, ages, as well as impact parameters and impact velocities, location of collision within the disk, etc. In turn, implementing a more realistic treatment of collisions in population synthesis models, e.g., by using scaling laws inferred  from SPH simulations, will affect the impact conditions obtained from such models highlighting the interplay between SPH and disk simulations.  

Second, this work did not include pre-impact rotation. Investigating and quantifying the effect of pre-impact rotation is important since this effect could provide an additional stabilisation against gravitational contraction and affect the outcome of collisions. Since the impact simulations cover a time scale of hundreds of years, the heliocentric orbits of the clumps and their cooling as well as the effect of the disk on the clumps should also be considered. Similarly, the presence of magnetic fields could also affect the disk's evolution and the final properties of the clumps \citep{dengFormationIntermediatemassPlanets2021,kubliCharacterizingFragmentationSubJovian2023}. 
\par 

Third, our collision simulations do not account for the tidal interaction with the central star. As a result, clump masses inferred in our simulations correspond to an upper limit. The exact final mass of the merger remnants could be estimated using the Hill radius \citep{hamiltonOrbitalStabilityZones1992} and would depend on the stellar mass and semi-major axis. Future studies could include the influence of the central star by directly adding its gravitational field during the collisions.

Fourth, it would be desirable to further investigate the shock-induced dynamical collapse and assess its significance in the DI model for giant planet formation. This could lead to a possible stimulated formation pathway along the same lines of shock-induced star formation due to collisions of molecular clouds or galaxies \citep{scovilleHighMassStarFormation1986,jogTriggeringMechanismEnhanced1992}.

Finally, future research could focus on fly-by events with the closest separation being larger than $R_{crit}$ to further investigate pure tidal interactions between the clumps and to determine whether continuous fly-bys have the ability to critically shape the body, a possible mechanism inspired by galactic dynamics \citep{mooreGalaxyHarassmentEvolution1996}.

\section{Summary and conclusions} \label{sec:conclusions}
We investigated the outcome of collisions between planetary clumps formed by DI using the SPH code \texttt{pkdgrav3} and the SCvH EoS. We performed more than 250 simulations of 3D impacts with different initial conditions. We found a rich diversity of collision outcomes and captured the transition area between them. 
The main conclusions of our study can be summarised as follows:

\begin{itemize}

    \item Clumps are prone to erosion and disruption.  Perfect merging as assumed in population synthesis models is very rare. Only for very grazing and low velocity collisions between young or mid clumps this assumption can be justified.

    \item Hit-and-run collisions typically lead to moderate mass loss but can have a significant effect on the post-impact density and temperature profile and therefore the survivability of the clumps. 
    
    \item Disruptive collisions can occur at rather low relative velocities of a few \SI{}{\kilo\meter\per\second}. Younger clumps are more vulnerable to disruption. 

    \item Dynamical collapse can be accelerated by a collision. This increases the survival probability of the remnant clump (proto-planet) within the proto-planetary disk. 

    \item Initiating the dynamical collapse with a collision is common for evolved clumps but also possible for clumps in the middle of their pre-collapse phase and impact velocities as low as a few hundred \SI{}{\meter\per\second}.  

\end{itemize}

Overall, our study clearly shows that the perfect merging assumption is not justified. Population synthesis models should account for the diversity in collision outcomes which in turn will have a significant impact on the inferred population of giant planets formed via disk instability. 

\begin{acknowledgements}
We thank the anonymous referee for valuable suggestions and comments that helped to improve the paper. RH acknowledges support from SNSF grant $200020\_215634$. Part of this work has been carried out within the framework of the National Centre of Competence in Research PlanetS supported by the Swiss National Science Foundation under grants 51NF40\_182901 and 51NF40\_205606. We acknowledge access to Piz Daint and Eiger@Alps at the Swiss National Supercomputing Centre, Switzerland under the University of Zurich's share with the project ID UZH4. We also thank Mirco Bussmann, Oliver Schib, Lucio Mayer, Noah Kubli, Christoph Mordasini and Saburo Howard for valuable discussions and suggestions.
\end{acknowledgements} 

\section*{Data availability} \label{sec:data}

The data that support the findings of this study will be shared on reasonable request to the corresponding author.

\bibliographystyle{aa}
\bibliography{references}

\begin{appendix}\label{sec:appendix}
\section{Initial conditions}

\begin{figure*}
\centering
\includegraphics[width=\hsize]{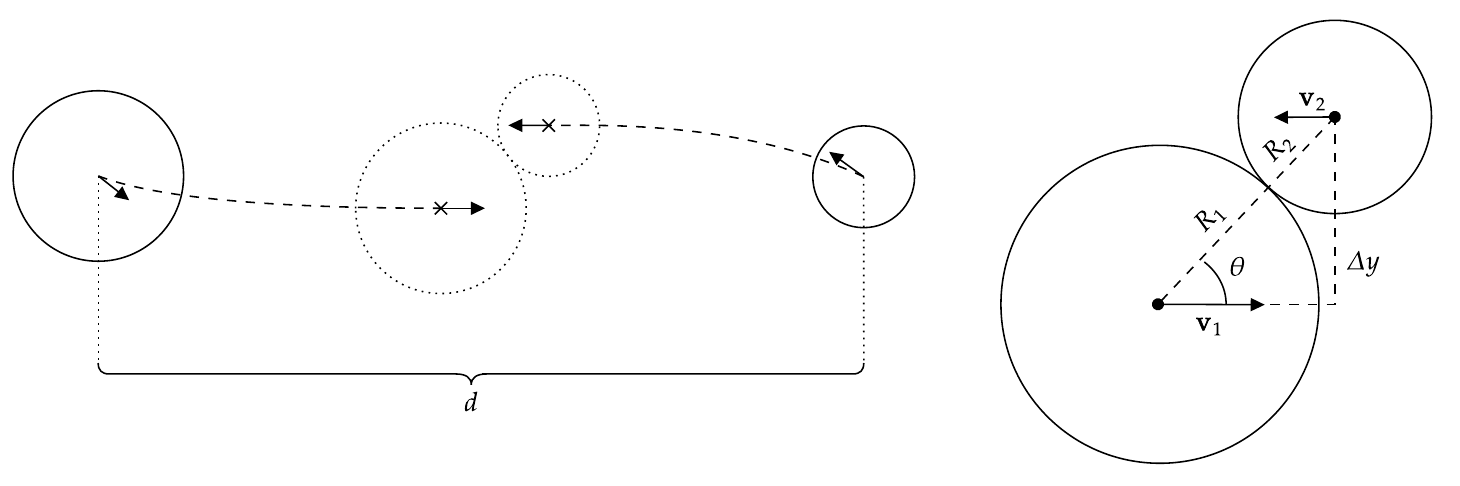}
\caption{Left: In order to respect the impact conditions while also having an initial separation between the clumps, we use a back integration scheme for their path to collision. Given an impact parameter $b$ and velocity $v_{imp} = |\textbf{v}_1-\textbf{v}_2|$, the trajectory of the clumps has been integrated back in time from their collision to the wished separation of $d = 5R_{crit}$. The integration scheme is performed within the point-mass approximation. Right: Geometrical parameters of impact conditions.}
\label{fig:ic}
\end{figure*}

\begin{table*}
\caption{Summary of all simulations with their initial conditions.\label{tab:sim_database}}
\centering
\begin{tabular}{l|ll}
\hline\hline
Colliding clumps & $v_{imp}/v_{esc}$ & $b$ \\
\hline
$3 M_J$ young -- $3 M_J$ young & 0.7 & \{0, 0.1, 0.2, 0.3, 0.4, 0.5, 0.7\} \\
 & 0.8 & \{0, 0.5, 0.7\} \\
 & 0.85 & \{0.5\} \\
 & 0.9 & \{0, 0.5, 0.7\} \\
 & 1.05 & \{0, 0.1, 0.2, 0.3, 0.4, 0.5, 0.6, 0.7\}  \\
 & 1.5 &  \{0, 0.1, 0.2, 0.3, 0.4, 0.5, 0.6, 0.7\} \\
 & 2  & \{0, 0.1, 0.2, 0.3, 0.4, 0.5, 0.6, 0.7\}\\ \hline
$5 M_J$ evolved -- $5 M_J$ evolved & 1.05 & \{0, 0.3, 0.4, 0.45, 0.5, 0.6, 0.65, 0.7\}  \\
 & 1.5 &  \{0, 0.1, 0.2, 0.3, 0.4, 0.5, 0.6, 0.7\} \\
 & 2  & \{0, 0.1, 0.2, 0.3, 0.4, 0.5, 0.6, 0.7\}\\ \hline
$10 M_J$ evolved -- $1 M_J$ evolved & 1.05 & \{0, 0.3, 0.5, 0.6, 0.62, 0.64, 0.65, 0.7\}  \\
 & 1.5 &  \{0, 0.3, 0.5, 0.6, 0.7, 0.8\} \\
& 2  & \{0, 0.1, 0.2, 0.3, 0.4, 0.5, 0.6\}\\ \hline
$10 M_J$ evolved -- $3 M_J$ young & 1.05 & \{0, 0.1, 0.2, 0.3, 0.4, 0.5, 0.6, 0.7\}  \\
& 1.5 &  \{0, 0.1, 0.2, 0.3, 0.4, 0.5, 0.6, 0.7\} \\
 & 2  & \{0, 0.1, 0.2, 0.3, 0.4, 0.5, 0.6, 0.7\}\\ \hline
$10 M_J$ evolved -- $5 M_J$ evolved & 1.05 & \{0, 0.3, 0.5, 0.52, 0.55, 0.58, 0.6, 0.65, 0.7\} \\
 & 1.5 &  \{0, 0.1, 0.2, 0.3, 0.4, 0.5, 0.6, 0.7\} \\
 & 2  & \{0, 0.1, 0.2, 0.3, 0.4, 0.5, 0.6, 0.7\}\\ \hline
$10 M_J$ mid -- $5 M_J$ mid & 0.1 & \{0, 0.5, 0.7\} \\
 & 0.3 & \{0, 0.5, 0.7\} \\
 & 0.6 & \{0, 0.5, 0.7\} \\
 & 0.7 & \{0, 0.5, 0.7\} \\
 & 0.8 & \{0, 0.5, 0.7\} \\
 & 1.05 & \{0, 0.1, 0.2, 0.3, 0.4, 0.5, 0.6, 0.7\} \\
 & 1.5 &  \{0, 0.1, 0.2, 0.3, 0.4, 0.5, 0.6, 0.7\} \\
 & 2  & \{0, 0.1, 0.2, 0.3, 0.4, 0.5, 0.6, 0.7\}\\ \hline
$10 M_J$ mid -- $10 M_J$ mid & 0.1 & \{0, 0.5, 0.7\}\\
 & 0.3 & \{0, 0.5, 0.7\} \\
 & 0.6 & \{0, 0.5, 0.7\} \\
 & 0.7 & \{0, 0.5, 0.7\} \\
 & 0.8 & \{0, 0.5, 0.7\} \\
 & 1.05 & \{0, 0.3, 0.5, 0.55, 0.65, 0.7\} \\
 & 1.5 &  \{0, 0.3, 0.35, 0.45, 0.5, 0.7\} \\
 & 2  & \{0, 0.3, 0.37, 0.44, 0.5, 0.7\}\\ \hline
$10 M_J$ evolved -- $10 M_J$ evolved & 0.9 & \{0, 0.5, 0.7\} \\
 & 0.95 & \{0, 0.5\} \\
 & 1.05 & \{0, 0.3, 0.4, 0.5, 0.6, 0.65\} \\
 & 1.3 &  \{0.3, 0.35, 0.4, 0.45, 0.5, 0.55, 0.6, 0.65, 0.7\} \\
 & 1.5 &  \{0, 0.1, 0.2, 0.3, 0.4, 0.5, 0.6, 0.65, 0.7\} \\
 & 1.75 &  \{0, 0.1, 0.2, 0.3, 0.4, 0.5, 0.6, 0.65, 0.7\} \\
 & 2  & \{0, 0.1, 0.2, 0.3, 0.4, 0.5, 0.6, 0.65, 0.7\}\\ \hline
$10 M_J$ evolved -- $10 M_J$ evolved ($N=10^5$) & 1.05 & 0 \\
$10 M_J$ evolved -- $10 M_J$ evolved ($N=10^7$) & 1.05 & 0 \\
\end{tabular}
\tablefoot{List of all performed collision simulations. Each simulation is defined by the combination of colliding clumps, impact velocity and impact parameter. The impact velocity is normalized to the mutual escape velocity of the clumps. The two last rows are part of the resolution study.}
\end{table*}

\begin{table*}
\caption{Summary of the collision outcomes from all performed simulations}\label{tab:outcome_database}
\centering
\begin{tabular}{l|llllllll}
\hline\hline
Colliding clumps & $N_{simulation}$ & $N_{PM}$ & $N_{erosion}$ & $N_{disruption}$ & $N_{PHR}$ & $N_{EHR}$ & $N_{collapse}$ \\
\hline
$3 M_J$ young -- $3 M_J$ young & 38 & 11 & 13 & 6 & 5 & 3 & 0  \\
\\
$5 M_J$ evolved -- $5 M_J$ evolved & 24 & 1 & 5 & 2 & 6 & 3 & 7 \\
\\
$10 M_J$ evolved -- $1 M_J$ evolved & 21 & 0 & 0 & 0 & 6 & 0 & 15  \\
\\
$10 M_J$ evolved -- $3 M_J$ young & 20 & 0 & 8 & 0 & 1 & 0 & 11  \\
\\
$10 M_J$ evolved -- $5 M_J$ evolved & 25 & 1 & 0 & 0 & 7 & 0 & 17  \\
\\
$10 M_J$ mid -- $5 M_J$ mid & 41 & 8 & 9 & 4 & 7 & 2 & 11  \\
\\
$10 M_J$ mid -- $10 M_J$ mid & 33 & 6 & 4 & 1 & 8 & 2 & 12  \\
\\
$10 M_J$ evolved -- $10 M_J$ evolved & 49 & 2 & 9 & 3 & 15 & 5 & 15  \\
\end{tabular}
\tablefoot{List of pair of colliding clumps with the number of simulations run and the occurrence of each type of outcome, respectively perfect mergers, erosions, disruptions, perfect hit-and-runs, erosive hit-and-runs and collisions resulting in dynamical collapse. }
\end{table*}

\end{appendix}

\end{document}